\documentclass{article}
\usepackage{spconf,amsmath,graphicx,amssymb}
\usepackage{array}
\usepackage{hyperref}
\newcolumntype{P}[1]{>{\centering\arraybackslash}p{#1}}
\newcolumntype{M}[1]{>{\centering\arraybackslash}m{#1}}

\newcommand{\bslname}{\textsc{SpeechSplit }}
\newcommand{\bslnamens}{\textsc{SpeechSplit}}
\newcommand{\ourname}{\textsc{SpeechSplit2.0 }}
\newcommand{\ournamens}{\textsc{SpeechSplit2.0}}

\title{\ournamens: Unsupervised speech disentanglement for voice conversion Without tuning autoencoder Bottlenecks}
\name{Chak Ho Chan$^{\star}$ \qquad Kaizhi Qian$^{\dagger}$ \qquad Yang Zhang$^{\dagger}$ \qquad Mark Hasegawa-Johnson$^{\star}$}
  
  \address{
    $^{\star}$ University of Illinois at Urbana-Champaign\\
    $^{\dagger}$MIT-IBM Watson AI Lab}

\begin{document}
%
\maketitle
\begin{abstract}


\bslname can perform aspect-specific voice conversion by disentangling     speech into content, rhythm, pitch, and timbre using multiple autoencoders in an unsupervised manner. However, \bslname requires careful tuning of the autoencoder bottlenecks, which can be time-consuming and less robust. This paper proposes \ournamens, which constrains the information flow of the speech component to be disentangled on the autoencoder input using efficient signal processing methods instead of bottleneck tuning. Evaluation results show that \ourname achieves comparable performance to \bslname in speech disentanglement and superior robustness to the bottleneck size variations. Our code is available at \url{https://github.com/biggytruck/SpeechSplit2}.

\end{abstract}
\section{Introduction}
\label{sec:intro}

Human speech conveys a rich stream of information, which may contain many components entangled with each other such as content, rhythm, pitch, timbre, emotion, and accent, etc. However, most speech applications only focus on a narrow subset of the many components. For example, automatic speech recognition only focuses on the content; speaker recognition and emotion recognition focus on timbre and emotion respectively; voice conversion focuses mainly on timbre. Based on this, it is beneficial to disentangle the speech components of interest for a variety of speech applications, such as speech recognition \cite{disentangle-for-ASR}, speech synthesis \cite{disentangle-for-synthesis1}, emotion analysis \cite{VAW-GAN}, privacy protection \cite{disentangle-for-privacy2}, and voice conversion \cite{AutoVC}.

This paper focuses on speech disentanglement for voice conversion. Voice conversion aims to modify the voice characteristics of speech while keeping the linguistic content unchanged, which is an area where speech disentanglement has been frequently explored. The majority of the voice conversion systems focus on timbre conversion, where content and timbre disentanglement is the key to success. As an early attempt, VAE-VC \cite{hsu2016voice} directly applies a variational autoencoder (VAE) for timbre disentanglement. Later, Chou et al. \cite{Chou2018MultitargetVC} and ACVAE-VC \cite{kameoka2018acvae} disentangle timbre using an auxiliary speaker classifier. Inspired by image style transfer, StarGAN-VC \cite{kaneko2019stargan}, CycleGAN-VC \cite{Kaneko2018CycleGANVCNV} adapted StarGAN \cite{Choi2018StarGANUG} and CycleGAN \cite{Zhu2017UnpairedIT} respectively for voice conversion. AutoVC \cite{AutoVC} disentangles speakers and content by directly tuning the bottleneck dimensions of a vanilla autoencoder. The following AutoVC-F0 \cite{qian2020f0} improves pitch disentanglement by conditioning on pitch contour. AutoPST \cite{Qian2021GlobalPS} further disentangles rhythm using similarity-based time resampling. To improve the granularity in speech disentanglement, \bslname \cite{SpeechSplit} disentangles speech into content, rhythm, pitch, and timbre using three encoders with carefully tuned bottlenecks. Although \bslname is effective for rhythm, pitch, and timbre conversion, it has two problems. First, bottleneck tuning is time- and resource-consuming. Second, re-tuning is required for a different dataset.

In this paper, we propose \ournamens, in which we apply efficient signal processing techniques to alleviate the laborious bottleneck tuning without modifying the network architecture of \bslnamens. We show that by processing the encoder inputs, we can control the information flowing into the model so that it can learn a disentangled representation for each component with reduced demands for bottleneck tuning. Experiments show that the disentangling performance of the proposed method remains robust under different bottleneck dimensions without tuning the bottleneck dimensions. It is worth mentioning that similar to \bslnamens, \ourname achieves speech disentanglement in an unsupervised manner by only training on reconstruction loss, which does not require parallel data or text transcriptions.

\section{Method}
\label{sec:method}

\subsection{\bslnamens}

\begin{figure}[t]
  \centering
  \includegraphics[scale=0.2]{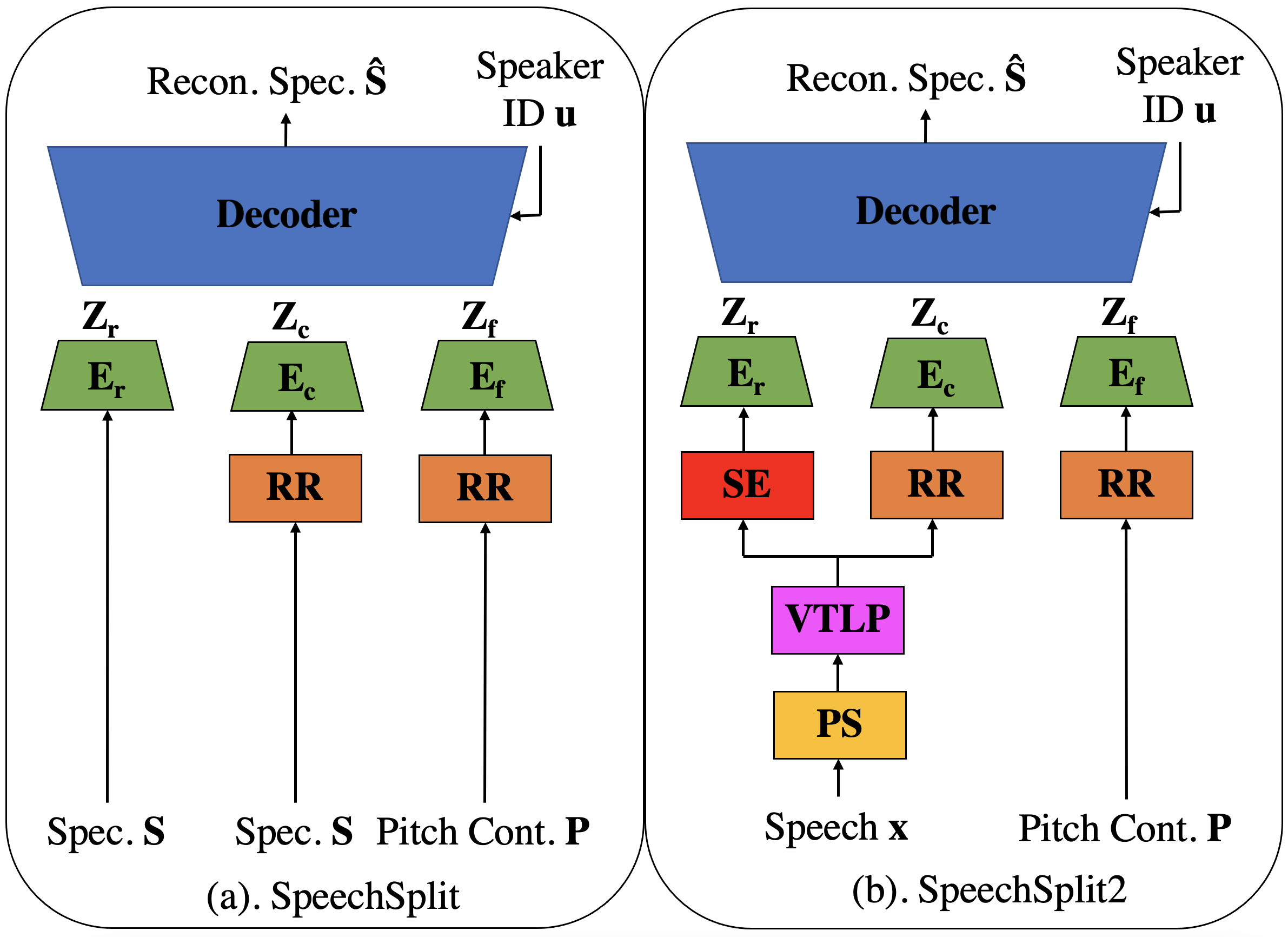}
  \caption{(a).\ \bslname (b).\ \ournamens. `RR' denotes random resampling, `PS' denotes pitch smoother, `VTLP' denotes vocal tract length perturbation, and `SE' denotes extracting spectral envelope by ceptral liftering}
  \label{fig:SpeechSplit}
\end{figure}

\bslname\cite{SpeechSplit} is an autoencoder-based generative model that decomposes speech into four components: rhythm, content, pitch, and timbre. Three encoders, denoted as $E_r$, $E_c$, and $E_f$, are used to encode rhythm, content, and pitch information respectively and can be formulated as:
\begin{equation}
  \mathbf{Z}_r = E_r(\mathbf{S}), \mathbf{Z}_c = E_c(R(\mathbf{S})), \mathbf{Z}_f = E_f(R(\mathbf{P}))
  \label{eq1}
\end{equation}
where $\mathbf{S}=[\mathbf{s}_1,...,\mathbf{s}_T]^{T}$ is the mel-spectrogram, $\mathbf{P}=[\mathbf{p}_1,...,\mathbf{p}_T]^{T}$ is the one-hot representation of the quantized pitch contour that is normalized to have the same mean and variance with respect to each speaker, $R$ denotes random resampling operation along time, and $\mathbf{Z}_r$, $\mathbf{Z}_c$, $\mathbf{Z}_f$ are the encoder outputs. The decoder $D$ then takes all three encoded representations and a speaker embedding $\mathbf{u}$ (a one-hot embedding, or an embedding computed by an automatic speaker ID network) to generate an output spectrogram $\mathbf{\hat{S}}$:
\begin{equation}
  \mathbf{\hat{S}} = D(\mathbf{Z}_r, \mathbf{Z}_c, \mathbf{Z}_f, \mathbf{u})
  \label{eq2}
\end{equation}
The model is trained to reconstruct the input spectrogram:
\begin{equation}
  \theta=\min_{\theta}\mathop{\mathbb{E}}[||\mathbf{\hat{S}}-\mathbf{S}||^{2}_{2}]
  \label{eq3}
\end{equation}
where $\theta$ denotes the model parameters. The model architecture is shown in Fig.~\ref{fig:SpeechSplit}(a). Since the random resampling operation corrupts the rhythm in the input to $E_c$ and $E_f$, only $E_r$ can observe the complete rhythm information. Thus, we can carefully tune the bottleneck dimension of $E_r$ so that it only encodes the rhythm representation. Similarly, given that $E_r$ only encodes the rhythm, we can carefully tune the dimension of $E_c$ so that it only encodes the content information. Finally, the pitch contour still contains certain timbre information even after being randomly resampled, but it is assumed in \cite{SpeechSplit} that normalization can remove the timbre, so $E_f$ can have a relatively large bottleneck. Nevertheless, tuning the entire network is laborious, considerably limiting the model's applicability on different datasets and tasks.

\subsection{\ournamens}

The proposed algorithm, \ournamens, addresses the bottleneck tuning issue by using efficient signal processing algorithms at the front end of the encoders to corrupt the component to be disentangled, which significantly reduce the sensitivity of the model to the bottleneck dimensions. Fig.~\ref{fig:SpeechSplit}(b) shows the model architecture.

\subsubsection{Content Encoder Input}

To remove the pitch information from an utterance $\mathbf{x}$, we implement a \textit{pitch smoother}, which first analyzes the signal using the WORLD vocoder \cite{WORLD}:
\begin{equation}
  \mathbf{f}, \mathbf{a}, \mathbf{E} =A_{WORLD}(\mathbf{x})
  \label{eq4}
\end{equation}
where $\mathbf{f}$, $\mathbf{a}$, and $\mathbf{E}$ denote F0, aperiodicity, and spectral envelope extracted by the WORLD analyzer $A_{WORLD}$. We then replace all the voiced frames in $\mathbf{f}$ with its own voiced mean to obtain a \textit{normalized F0 contour} $\mathbf{\bar{f}}$ and re-synthesize the signal using the WORLD synthesizer $S_{WORLD}$:

%
\begin{equation}
  \mathbf{\bar{x}} = S_{WORLD}(\mathbf{\bar{f}}, \mathbf{a}, \mathbf{E})
  \label{eq6}
\end{equation}
where we denote $\mathbf{\bar{x}}$ as the \textit{monotonic utterance} with a corresponding \textit{monotonic spectrogram} $\mathbf{\bar{S}}$. Since all the elements in the F0 contour are replaced by its mean, the intonation dynamics along time are removed. Fig.~\ref{fig:model_input}(a) and ~\ref{fig:model_input}(b) show the spectrogram and monotonic spectrogram for utterance \textit{``Please call Stella.''} Note that the tone movements along time in the monotonic spectrogram, e.g., the dropping tone of \textit{``Stella''}, are completely flattened, indicating the pitch dynamics are discarded. Next, to corrupt the timbre information, we use Vocal Tract Length Perturbation (VTLP) \cite{vtlp}, which modifies the timbre by warping the frequency and has been extensively used in various tasks \cite{vtlp-for-dsr, vtlp-for-tts}. Formally,
\begin{equation}
  \mathbf{\widetilde{x}} = H(\mathbf{\bar{x}}, \alpha)
  \label{eq7}
\end{equation}
where $H$ denotes the VTLP operation, $\alpha \sim U(0.9, 1.1)$ is a warping factor, and $\mathbf{\widetilde{x}}$ is referred to as the \textit{perturbed utterance} with a corresponding \textit{perturbed spectrogram} $\mathbf{\widetilde{S}}$. We perturb each training utterance with a random $\alpha$ to prevent the content encoder from easily recovering the timbre information. Compared to Fig.~\ref{fig:model_input}(b), the perturbed spectrogram in Fig.~\ref{fig:model_input}(c) further shifts the formant frequencies down, resulting in a deeper voice. Finally, we randomly resample $\mathbf{\widetilde{S}}$ to corrupt the rhythm information to obtain the content encoder input $\mathbf{S}_c$ following the same procedure as in \bslnamens.


\subsubsection{Rhythm Encoder Input}

\bslname hypothesized a \textit{``fill in the blank''} mechanism: the rhythm representation provides blanks corresponding to all the syllables and pauses in an utterance, and the decoder fills in these blanks with the respective content and pitch code. Based on this, we suggest that a good rhythm representation should (1) preserve very little about the content, pitch or timbre and (2) include some "indexing cues" to inform the decoder which part of the content and pitch representations should be filled in each blank. Thus, we propose to use a spectral envelope obtained by liftering the real cepstrum of the perturbed utterance $\mathbf{\widetilde{x}}$ as the input:
\begin{equation}
  \mathbf{\widetilde{C}} = DFT^{-1}(\log(|\mathbf{\widetilde{Y}}|))
  \label{eq9}
\end{equation}
\begin{equation}
  \mathbf{\widetilde{C}}_{l} = \mathbf{\widetilde{C}}\mathbf{L}
  \label{eq10}
\end{equation}
\begin{equation}
  \mathbf{S}_{r} = R(\exp(DFT(\mathbf{\widetilde{C}}_{l})))
  \label{eq11}
\end{equation}
where $\mathbf{\widetilde{Y}}=[\mathbf{\widetilde{y}}_1,...,\mathbf{\widetilde{y}}_T]^{T}$ is the spectrum of $\mathbf{\widetilde{x}}$, $\mathbf{\widetilde{C}}=[\mathbf{c}_1,...,\mathbf{c}_T]^{T}$ is the real cepstrum, $\mathbf{\widetilde{C}}_{l}$ is the liftered real cepstrum and $\mathbf{L}$ is a diagonalized low-quefrency lifter:
\begin{equation}
    \mathbf{L}_{ij}= 
    \mbox{diag}\left(0.5 u[n_c-i]+0.5u[n_c-i-1]\right),
    \label{eq12}
\end{equation}
where $u[\cdot]$ is the unit step function.
\begin{figure}[t]
  \centering
  \includegraphics[scale=0.56]{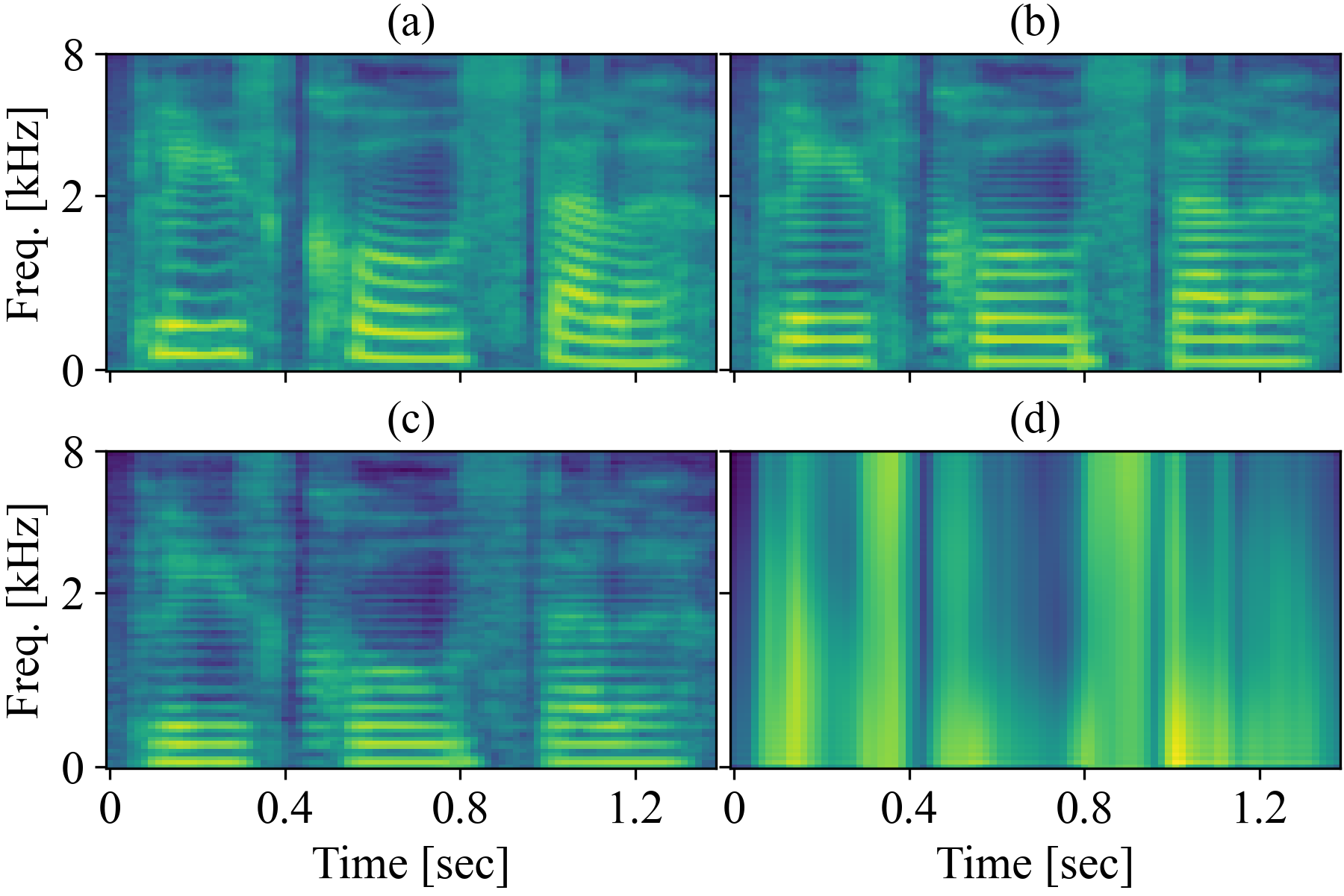}
  \caption{(a).\ Spectrogram for utterance \textit{``Please call Stella''} (b).\ Monotonic spectrogram (c).\ Perturbed spectrogram (d).\ Perturbed spectral envelope}
  \label{fig:model_input}
\end{figure}
If $n_c$ is low, then $\mathbf{S}_r$ will contain very little content, pitch, and timbre information since it is obtained from the perturbed utterance and the liftering operation discards most of the fine details. However, the rhythm information is preserved, because the remaining spectral features are distinct for different phonemes, from which the encoder can extract the rhythm information. Fig.~\ref{fig:model_input}(d) shows such a spectral envelope with $n_c=3$. As we can see, the phoneme variations are smoothed out, but the patterns are unique among different syllables and pauses.

\subsubsection{Pitch Encoder Input}

\bslname assumes that by normalizing the pitch contour and randomly resampling it in time, we can discard all other information but still preserve the pitch. Thus, we do not need to further process the input to the pitch encoder. 



However, \bslname requires a pitch converter to restore the temporal mismatch between the source speech and the target pitch contour to perform pitch conversion. The pitch converter is a smaller variant of the model, which consists of only a rhythm and a pitch encoder. Since the rhythm encoder takes in the original spectrogram as its input, it is able to learn how to temporally realign the pitch contour. In contrast, \ourname uses the spectral envelope as the rhythm input, which has insufficient information to realign the pitch contour. We fix the information gap by concatenating the perturbed spectrogram $\mathbf{\widetilde{S}}=[\mathbf{\widetilde{s}}_1,...,\mathbf{\widetilde{s}}_T]^{T}$ and the one-hot quantized pitch contour $\mathbf{P}=[\mathbf{p}_1,...,\mathbf{p}_T]^{T}$:
\begin{equation}
  \mathbf{S}_p = R([\begin{bmatrix} \mathbf{\widetilde{s}}_{1} \\ \mathbf{p}_{1} \end{bmatrix} ,...,\begin{bmatrix} \mathbf{\widetilde{s}}_{T} \\ \mathbf{p}_{T} \end{bmatrix}]^{T})
  \label{eq13}
\end{equation}
%
where $\mathbf{S}_p$ is the pitch input in the pitch converter. By concatenating $\mathbf{\widetilde{S}}$ and $\mathbf{P}$, the perturbed spectrogram can bridge the information gap between the pitch and the rhythm encoder. During conversion, the decoder can first extract the alignment information from the coarse spectral envelope and the fine-grained spectrogram, and then adjust the pitch contour length accordingly.

\begin{figure}[t]
  \centering
  \includegraphics[scale=0.4]{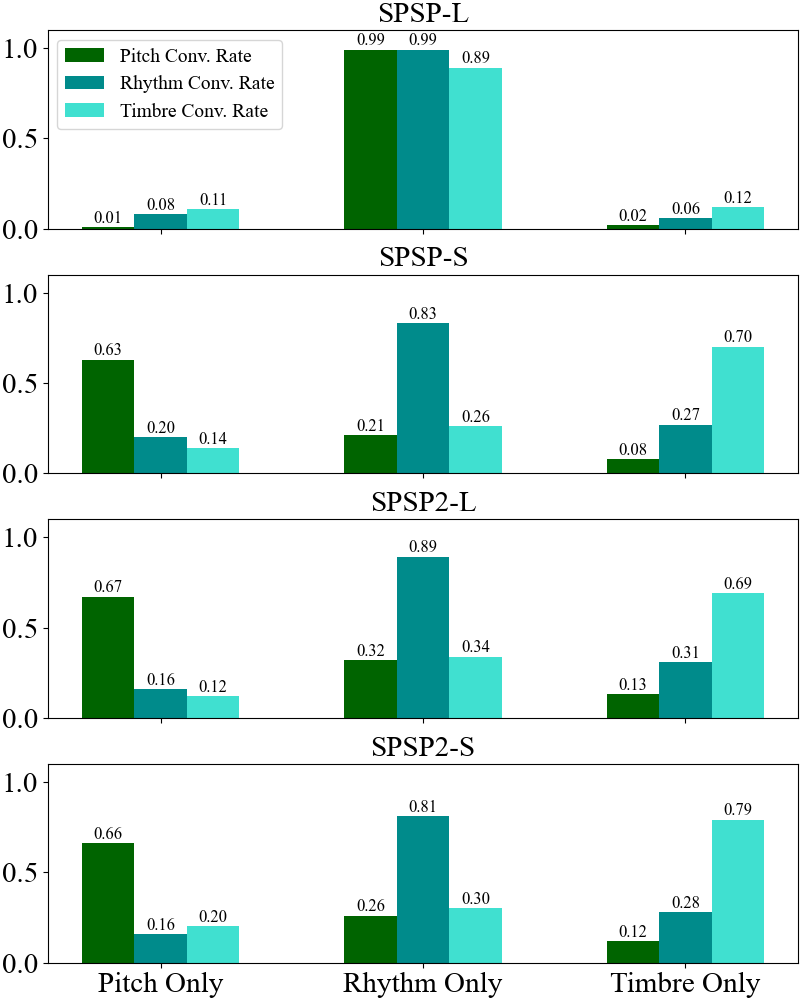}
  \caption{Conversion rates of different models}
  \label{fig:mturk}
\end{figure}

\section{Experiments}
\label{sec:experiments}


We evaluate the results on four models: two small models, SPSP-S and SPSP2-S, which are \bslname and \ourname whose bottleneck configurations are carefully tuned to have the right dimension for the VCTK corpus~\cite{vctk}; two large models, SPSP-L and SPSP2-L, which have much wider bottlenecks by increasing the encoder dimensions and decreasing the downsampling factors. Since SPSP-L's bottleneck is too wide, the rhythm encoder can encode all four aspects. Consequently, the model will only reconstruct the signal that provides the rhythm representation and fail on all types of conversion. The SPSP2 encoders, by contrast, should be robust to different bottleneck settings so that SPSP2-L and SPSP2-S have comparable speech conversion ability as the carefully tuned SPSP-S. The model configurations are listed in Table 1, in which Dim\textsubscript{R}, Dim\textsubscript{C}, and Dim\textsubscript{F} denotes the dimension for the rhythm, content and pitch encoder and DS\textsubscript{R}, DS\textsubscript{C}, and DS\textsubscript{F} denotes the respective downsampling factor. We use $n_c=3$ for the liftering operation. All other configurations are the same as in \bslname \cite{SpeechSplit}. Audio samples are available at \url{https://biggytruck.github.io/spsp2-demo}.

\begin{table}
  \label{tab:model configurations}
  \centering
  \resizebox{0.97\linewidth}{!}{%
  \begin{tabular}{c|c|c|c|c|c|c} 
     \hline\hline
     Model & $\text{Dim}_{R}$ & $\text{Dim}_{C}$ & $\text{Dim}_{F}$ & $\text{DS}_{R}$ & $\text{DS}_{C}$ &
     $\text{DS}_{F}$ \\ 
     \hline
    SPSP-L & 32 & 32 & 32 & 1 & 1 & 1 \\
    SPSP-S & 1 & 8 & 32 & 8 & 8 & 8 \\ 
    SPSP2-L & 32 & 32 & 32 & 1 & 1 & 1 \\ 
    SPSP2-S & 1 & 8 & 32 & 8 & 8 & 8 \\ 
     \hline\hline
     \end{tabular}}
    \caption{Model configurations}
\end{table}

\subsection{Subjective Evaluations}

We first evaluate the models' performance on \textit{Amazon Mechanical Turk}. For each model, we apply pitch-only, rhythm-only, and timbre-only conversion on 20 utterance pairs that are perceptually distinct in the aspect of conversion and present the results to five subjects. Each subject is asked to listen to a source and a target reference in random order and determine which reference the converted speech is more similar to in terms of all three aspects. We then compute the \textit{conversion rate} for each aspect, defined as the percentage of answers selecting the target reference, as shown in Fig.~\ref{fig:mturk}. For SPSP-L, all three conversion rates are very high for rhythm-only conversion but low for pitch-only and timbre-only conversion, indicating the rhythm encoder encodes all the information while the others only encode very trivial information. SPSP-S and both variants of SPSP2 successfully convert the aspect of interest without modifying the other attributes of the source utterance, which suggests that SPSP2 has the similar conversion capability as a carefully fine-tuned SPSP regardless of the bottleneck dimensions.

\begin{figure}[t]
  \centering
  \includegraphics[scale=0.4]{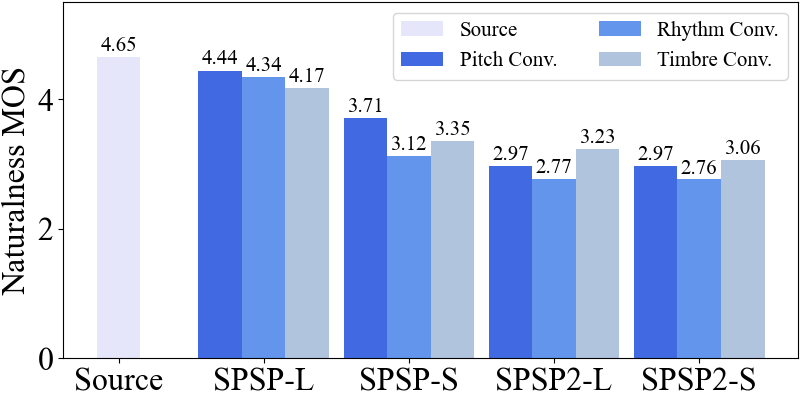}
  \caption{Naturalness evaluation on different models}
  \label{fig:naturalness}
\end{figure}

For all the converted utterances, we also ask the listeners to evaluate the naturalness with a 5-scale mean opinion score (MOS), as shown in Fig.~\ref{fig:naturalness}. SPSP-L achieves the highest MOS since it essentially performs reconstruction of the input from the rhythm code. Compared to the fine-tuned SPSP-S, both variants of SPSP2 show some degree of degradation in naturalness. 
We hypothesize that the pitch smoother introduces artifacts in the smoothed audio mainly due to pitch tracking errors, which ultimately affect the naturalness.  

\subsection{Objective Evaluation}

For objective evaluation, we are mainly interested in the intelligibility of the converted speech. Thus, we measure the character error rate (CER) of the transformed utterances using Google Cloud ASR, as shown in Table 2. Similar to the naturalness measurement, the artifacts result in higher CER in SPSP2 as compared to SPSP. It is worth noting that SPSP2-L has lower CER than SPSP2-S in all three types of conversion, suggesting a larger bottleneck preserves more content information.
Given the evaluation results for naturalness and intelligibility, we may in the future improve SPSP2 by (1) replacing the pitch contour with better pitch representation and 
(2) designing more subtle bottleneck mechanisms to better preserve aspect-specific information with minimum content loss.

\begin{table}
  
  \label{tab:CER}
  \centering
  \begin{tabular}{M{1.3cm}|M{1.3cm}|M{1.3cm}|M{1.3cm}}
    \hline\hline
    Model & Pitch Conv. & Rhythm Conv. & Timbre Conv. \\  \hline
    SPSP-L & 12.9\% & 14.4\% & 9.8\% \\ 
    SPSP-S & 30.8\% & 46.3\% & 34.0\% \\ 
    SPSP2-L & 37.8\% & 54.5\% & 39.2\% \\ 
    SPSP2-S & 54.4\% & 62.6\% & 43.5\% \\ 
    \hline\hline
  \end{tabular}
  \caption{CER of converted utterances by model}
\end{table}

\section{Conclusion}

\label{sec:conclusion}
The proposed model, \ournamens, demonstrates equivalent capability in speech disentanglement as \bslname without the need to tune the bottleneck. We show that by modifying the inputs using efficient signal processing techniques, we can drive the encoders to learn a disentangled representation for different aspects of speech. In the future, we will focus on improving the naturalness and intelligibility of the generated speech and attempt to disentangle other components such as emotion and accent.


\bibliographystyle{IEEEbib}
\bibliography{strings,refs}

\end{document}